\documentclass{pasj00}

\newcommand{\Msun}{M_{\odot}}

\newcommand{\pc}{{\rm \, pc}}
\newcommand{\kpc}{{\rm \, kpc}}

\newcommand{\MHz}{{\rm \, MHz}}
\newcommand{\GHz}{{\rm \, GHz}}
\newcommand{\kmps}{{\rm \, km \, s^{-1}}}
\newcommand{\jkmps}{{\rm \, Jy \, km \, s^{-1}}}
\newcommand{\K}{{\rm \, K}}
\newcommand{\convf}{{\rm cm^{-2}\, [K \, km \, s^{-1}]^{-1}}}
\newcommand{\vsys}{V_{\rm sys}}

\newcommand{\Ha}{\rm H{\alpha}}

\newcommand{\Omegab}{\Omega_{\rm b}}
\newcommand{\Msunpcpc}{{\Msun \, \rm pc^{-2}}}
\newcommand{\yr}{{\rm \, yr}}

\begin{document}
\SetRunningHead{J. Koda and Y. Sofue}{Gas Dynamics and Star Formation in NGC 4303}
\Received{2005 December 27}
\Accepted{2006 February 13}

\title{The Virgo High-Resolution CO Survey: VI. \\
Gas Dynamics and Star Formation Along the Bar in NGC 4303}

\author{Jin \textsc{Koda}}
\affil{California Institute of Technology, MS 105-24, Pasadena, CA 91125, USA\\
and National Astronomical Observatory, Mitaka, Tokyo, 181-8588}
\email{koda@astro.caltech.edu}
\and
\author{Yoshiaki \textsc{Sofue}}
\affil{Institute of Astronomy, The University of Tokyo, Mitaka, Tokyo, 181-0016}
\email{sofue@ioa.s.u-tokyo.ac.jp}


%

\KeyWords{galaxies: individual(NGC 4303) -- galaxies: ISM -- galaxies: kinematics and dynamics} 

\maketitle

\begin{abstract}
We present CO interferometer observations of the barred galaxy NGC 4303 (M61).
This galaxy has a strong gas concentration at the central region and
prominent offset ridges at the leading sides of the bar. Sharp velocity
gradients are apparent across the ridges. The brightness temperature
in the ridges is low, indicating the existence of unresolved molecular clouds.
Additionally, an analysis of the newborn stellar clusters revealed
in {\it HST} images suggests that the parent molecular clouds had masses
of $10^{4-6}\Msun$.
The observed shear velocity gradient
across the ridges is too small to break up giant molecular clouds.
Therefore, the clouds are likely to survive passage through the ridges.
We discuss a cloud orbit model in a bar potential for understanding
the gas distribution,
dynamics, and star formation in NGC 4303. The model reproduces the narrow
offset ridges and sharp velocity gradients across the ridges, although
no shock is associated with the ridges.
We discuss cloud-cloud collisions (and close interactions)
as a possible triggering mechanism for star formation.
The newborn stellar clusters in NGC 4303 are located predominantly at the
leading sides of the offset ridges. Cloud orbits are densely populated in
the region where the clusters are found, suggesting a high collisional
frequency and possibly a high rate of triggered star formation.
Cloud-based dynamics is less dissipative than smooth hydrodynamic models,
possibly extending the timescales of gas dynamical evolution and
gas fueling to central regions in barred galaxies.

\end{abstract}

\section{Introduction}

Despite numerous observational and theoretical studies, the nature of
interstellar medium (ISM) gas dynamics remains uncertain \citep{com96}.
The dynamical response of the gas depends significantly on whether
the ISM is a continuous hydrodynamic fluid or a set of discrete
molecular clouds.
The hydrodynamic fluid produces shocks along spiral arms or
bars in galaxies, while the molecular clouds could ballistically pass
the spiral/bar structures without encountering a shock.
Such differences must change
not only the dynamical response of the gas (e.g. shock, dissipation),
but also the evolution timescale (e.g. gas inflow) and star formation.

The Galactic ISM is concentrated in discrete molecular clouds \citep{sco87},
and supports the dynamics of slightly collisional clouds.
Individual molecular clouds are rarely resolved in external galaxies;
however, there is a general belief that only continuous hydrodynamic
models could reproduce the narrow gas/dust lanes observed in barred galaxies,
since hydrodynamic fluids easily produce narrow shock lines along bars.
We demonstrate that an alternative scenario, cloud-based gas dynamics,
also reproduces the narrow gas distribution and associated
gas kinematics in bars.

The large-scale distribution of star-forming regions and young stars
provides a clue to understanding gas dynamics.
Star formation triggered by hydrodynamic shocks would be localized
in narrow shocked regions. Cloud-cloud interactions could potentially
occur everywhere in galactic disks and cause star formation.
We compare CO observations of the barred galaxy NGC 4303 to the
cloud-based gas dynamics model. The cloud-based model naturally explains
the distribution of young stellar clusters.

\begin{figure}
\begin{center}
\FigureFile(160mm,160mm){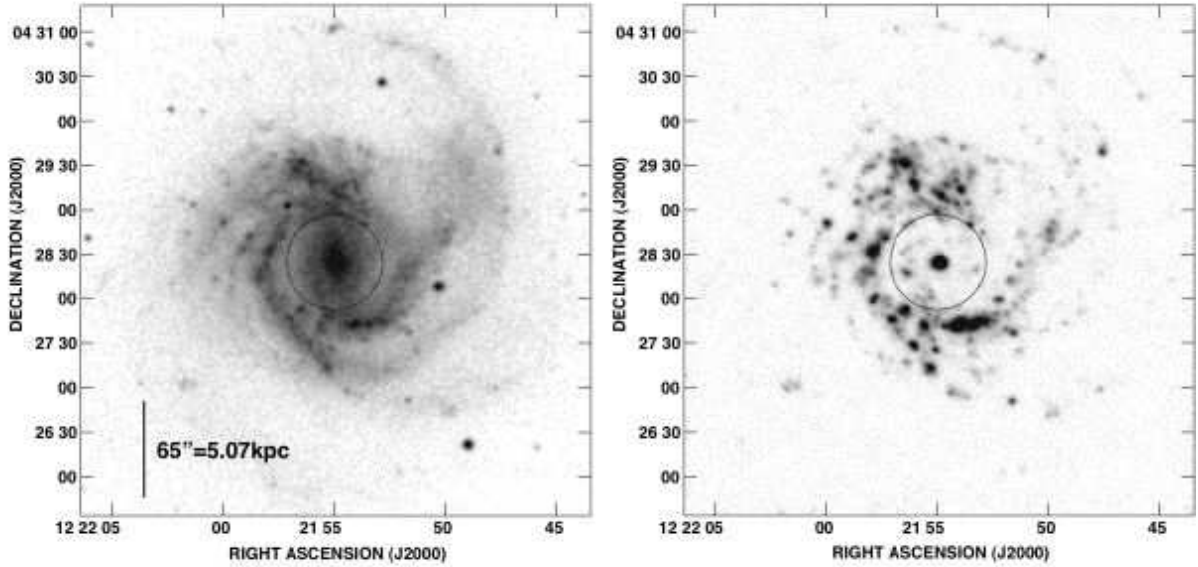}
\end{center}
\caption{
{\it Left: }$B$-band image of NGC 4303 from the Digitalized Sky Survey.
{\it Right:} $\Ha$ image from \citet{koo01}.
The circles represent the primary beam size of the $\rm ^{12}CO(J=1-0)$
observations ($65\arcsec$ HPBW).}
\label{fig:optimages}
\end{figure}

Figure \ref{fig:optimages} shows the $B$ and $\Ha$ band images of NGC4303;
most HII regions are associated with outer spiral arms, but almost no
apparent $\Ha$ emission is found in the bar. Similar trends have been
found in many other barred galaxies (\cite{dow96,reg96,gar96,she00}; 2002).
The absence of active star formation in bars is often attributed to
strong shear velocity in the bar \citep{rey98}. However, we point out below
that the shear in NGC 4303 is too small to break up dense
molecular clouds.

NGC 4303 hosts remarkable central activity: a nuclear secondary bar
\citep{col00}, prominent UV emission around the secondary bar
\citep{col97}, and an active galactic nucleus (AGN; \cite{col99}).
It has been a prime example of the starburst-AGN connection
\citep{col97,col99,col00,jim03}.
\citet{sch02} discussed gas fueling toward the nucleus due to
the secondary bar based on their CO data. Our focus, however,
is on the gas dynamics and star formation around the offset ridges
in the primary bar.

We describe the observations and data in \S \ref{sec:obs}, and results
in \S \ref{sec:res}. We introduce a cloud-orbit model in the bar
and show its ability to reproduce the observed CO distribution
and velocity structure (\S \ref{sec:mod}). We discuss star formation
along the bar using the cloud-based model (\S \ref{sec:sf}).
Conclusions
are summarized in \S \ref{sec:con}.
This paper is the sixth in a series on the Virgo CO survey at the
Nobeyama Millimeter Array (\cite{sof03}; 2003b; 2003c; \cite{ono04,nak05}).

\begin{center}
------ Figure 1 ------
\end{center}

\begin{center}
------ Table 1 ------
\end{center}

\section{Observational Data}\label{sec:obs}

\subsection{NMA CO(1-0) Observations}

We made aperture synthesis observations of the Virgo cluster galaxy
NGC 4303 in the CO $J=1-0$ line emission using the Nobeyama
Millimeter Array (NMA).
The observations were made between
1999 December and 2000 February with a single pointing center at
($\alpha_{1950}$, $\delta_{1950}$) = ($12^{\rm h}19^{\rm m}21^{\rm s}.6$,
$+4\arcdeg45\arcmin3\arcsec.0$). Six 10 m telescopes provide
the FWHP of about $65\arcsec$ at 115 GHz. We used three available
array configurations (AB, C, and D configurations).
The typical system noise temperature was about $400\K$ in single sideband.
Digital spectro-correlators \citep{oku00} have two spectroscopic
modes; we used the mode covering 512 MHz ($1331\kmps$)
with the 2 MHz ($5.2\kmps$) resolution. We observed the quasar 3C273
every 20 minutes for gain and bandpass calibrations. The absolute flux
of 3C273 was measured against Uranus and/or available flux-known quasars;
it was 14.2Jy for 1999 December, and 11.4Jy for 2000 January and February.
The uncertainties in the flux scale are about 20\%.

Raw visibility data were calibrated for passband response and complex
gain variation with the NRO/UVPROC-II package \citep{tsu97}.
The data were binned to 4 MHz resolution. Each channel data was
deconvolved with the CLEAN procedure using the NRAO/AIPS package.
We applied both natural and uniform weightings for the visibility
data and obtained $2\arcsec.4$- and $1\arcsec.3$-resolution maps,
respectively. We refer the $2\arcsec.4$-resolution map as ``map''.
The parameters of the observation and data cubes
are given in Table \ref{tab:obsparm} and \ref{tab:cbparam}, respectively.

\begin{center}
------ Table 2 ------
\end{center}

\begin{center}
------ Table 3 ------
\end{center}

Figures \ref{fig:comaps} shows zeroth- and first-moment maps, and
synthesized beam map. Figure \ref{fig:chmaps} and \ref{fig:obspv}
show the velocity channel maps and position-velocity (PV) diagrams,
respectively.
Primary beam attenuation is not corrected for these maps.
In the synthesized beam map (Figure \ref{fig:comaps}), prominent
sidelobes run in the north-south direction and are aligned
with the observed emission distribution.
These sidelobes made the deconvolution less obvious;
we arranged restricted CLEAN boxes and checked that the 2nd- and
3rd-order sidelobes are also reduced successfully in the deconvolution.
We estimated the recovered flux by comparing our data with
a single-dish observation at the FCRAO 14 m telescope \citep{ken88}.
Our map recovers 97\% of the total
flux in the central $45\arcsec$ (with the uncertainties of 20\%) .

\begin{center}
------ Figure 2 ------
\end{center}

\begin{center}
------ Figure 3 ------
\end{center}

\begin{center}
------ Figure 4 ------
\end{center}

\subsection{Supplied Optical/Near-IR Data}

We obtained optical and near-infrared images from the archives and literature
through the NASA/IPAC Extragalactic Database (NED): the $B$-band
image from the Digitized Sky Survey, $\Ha$-image (Figure \ref{fig:optimages}
and \ref{fig:optimages2}) from \citet{koo01}, and $K$-band image
(Figure \ref{fig:optimages2}) from \citet{moe01}.
We also obtained $HST$ images in the F450W and F814W-filters (P.I. Smartt)
through the HST archive at the Canadian Astronomy Data Center (CADC).
The F450W and F814W images are shown in Figure \ref{fig:hstglobal}.

The astrometry of the $B$- and $\Ha$-band images was determined using
the USNO-A2.0 catalog \citep{zac00}.
The $K$-band and $HST$ images do not contain enough cataloged stars;
we shift the central compact emission peak in each image to
the dynamical center
(\S \ref{sec:resoff}). The centroids of compact sources across these
images coincide with those in the lower-resolution $B$-band image.
The uncertainty of the absolute position in each image is about
$0\arcsec.5$.

\begin{center}
------ Figure \ref{fig:optimages2} ------
\end{center}

\begin{center}
------ Figure \ref{fig:hstglobal} ------
\end{center}

\section{Results}\label{sec:res}

Figure \ref{fig:comaps} shows two remarkable features in the gas
distribution:
(1) a central concentration within $r \sim 10 \arcsec$ ($780\pc$,
hereafter the central disk), and
(2) offset ridges extending from the central concentration
out to $r\sim30\arcsec$ ($2.3\kpc$).
The gas is rotating clockwise, assuming trailing spiral arms.
Below, we obtain the parameters of the gas disk and discuss the two
features.

\subsection{Global CO Properties}\label{sec:resglob}

Figure \ref{fig:prof} shows the radial profile of the CO integrated intensity
$I_{\rm CO} {\rm d}V \cos(i)$ and cumulative flux $S_{\rm CO}$.
Corrections for the primary beam response and inclination were applied.
We adopted the position angle ${\rm P.A.}= -45\arcdeg$, inclination
$i=30\arcdeg$ (see \S \ref{sec:rescent}), and the CO-to-$\rm H_2$ conversion
factor $X_{\rm CO}=1.8 \times 10^{20} \convf$ \citep{dam01}.
The surface density is almost constant at $\sim 500 \Msunpcpc$
from the center to $r\sim 4\arcsec$ ($310\pc$), and decreases
exponentially with a scale length of $\sim 4.5\arcsec$ ($350\pc$).
The total detected gas mass is $ 8.4 \times 10^8 \Msun$ within the radius
$13\arcsec$ ($1\kpc$), and is $1.0 \times 10^9\Msun$ within $30\arcsec$
($2.3\kpc$).

\begin{center}
------ Figure \ref{fig:prof}  ------ 
\end{center}

\subsection{Central Concentration}\label{sec:rescent}


The CO emission is very concentrated in the central disk ($r<10\arcsec$)
in Figure \ref{fig:comaps}. The velocity field is the typical
``spider diagram''. The gas is following almost pure circular rotation.
Spiral arms are curling along the outer edge of the central disk
(Figure \ref{fig:couni}). They are associated with slight perturbations
in the velocity field at the outer edges of the central disk.
We obtained kinematic parameters from the velocity-field assuming
pure circular rotation. The dynamical center, position angle, inclination,
and recession velocity were determined using the AIPS/GAL package
(Table \ref{tab:kinparm}). Since the perturbations
due to the spiral arms are small, the errors due to the perturbations
should be small.
The derived parameters are consistent with those from previous studies
\citep{col99, sch02}.

\begin{center}
------ Figure \ref{fig:couni}  ------ 
\end{center}

\subsection{Offset Ridges}\label{sec:resoff}

\begin{figure}
\begin{center}
\FigureFile(160mm,160mm){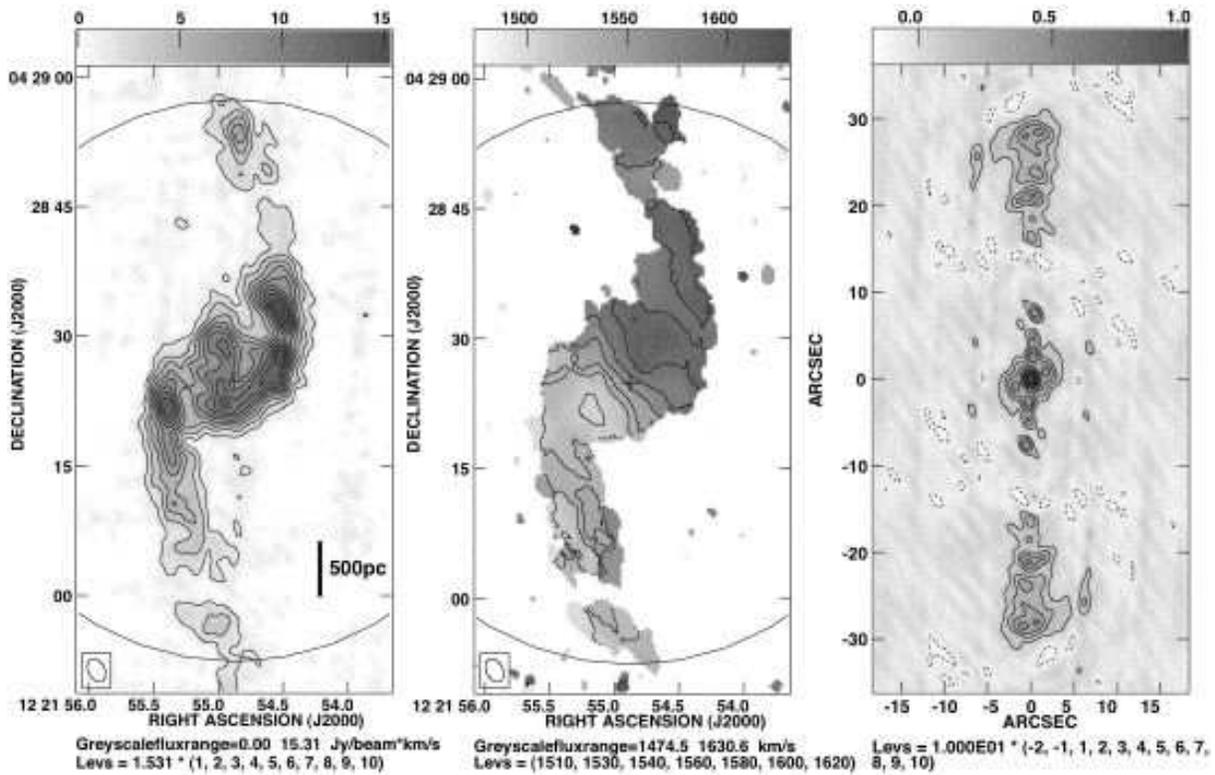}
\end{center}
\caption{
CO(1-0) zeroth- and first-moment maps of NGC 4303, and synthesized beam
map. A clip at the $2\sigma$ level was used to make these maps. The circles
represent the size of the primary beam. No primary beam
correction was applied.}
\label{fig:comaps}
\end{figure}

Two gas ridges run out from the central disk in the north and south
directions (Figure \ref{fig:comaps}). Figure \ref{fig:optimages2}
({\it left}) compares the gas distribution (grayscale; CO image) with
the stellar bar (contours; $K$-band image). The two ridges are at the
leading sides of the stellar bar. This type of ridge is referred
to as ``offset ridges'', and is often found in barred galaxies \citep{ish90,
sak99,she02}. The gas ridges coincide with dust lanes in
Figure \ref{fig:hstglobal}.
There are gas concentrations at the outermost ends of the detected
CO ridges (Figure \ref{fig:comaps}), which coincide with the stellar
bar ends. Such gas concentration is common in barred galaxies.
Note that no primary beam correction has been applied to this map;
the fluxes of these gas concentrations are therefore about 2-times
larger than the contour values shown in the figure.

\begin{figure}
\begin{center}
\FigureFile(160mm,160mm){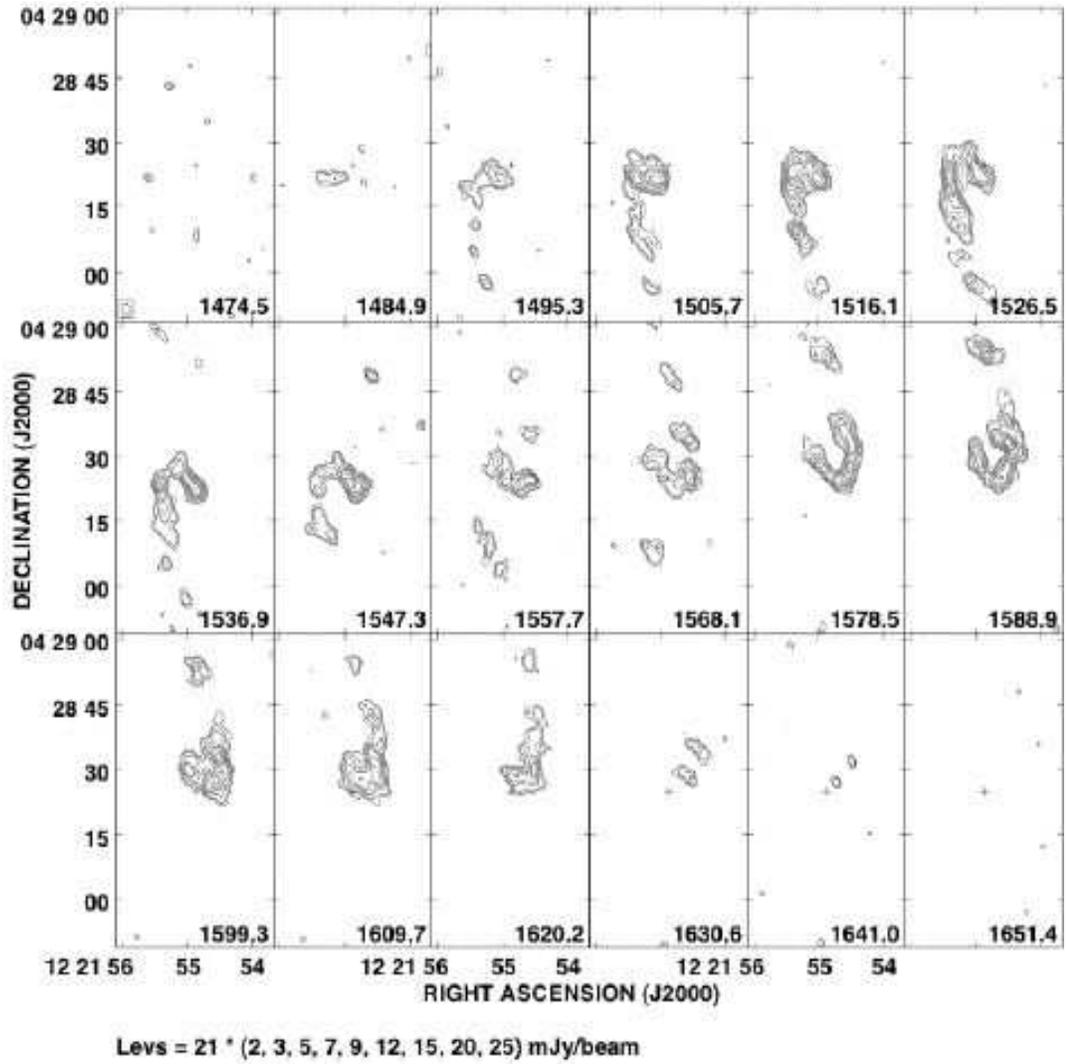}
\end{center}
\caption{
CO(1-0) velocity channel maps of NGC 4303 in the same region as in
Figure \ref{fig:comaps}. The velocities at the channels are shown at the
lower right corners. No primary beam correction was applied.
The $\rm 1\, mJy\,beam^{-1}$ corresponds to $0.017\rm\, K$ in
brightness temperature.}
\label{fig:chmaps}
\end{figure}

\begin{figure}
\begin{center}
\FigureFile(80mm,80mm){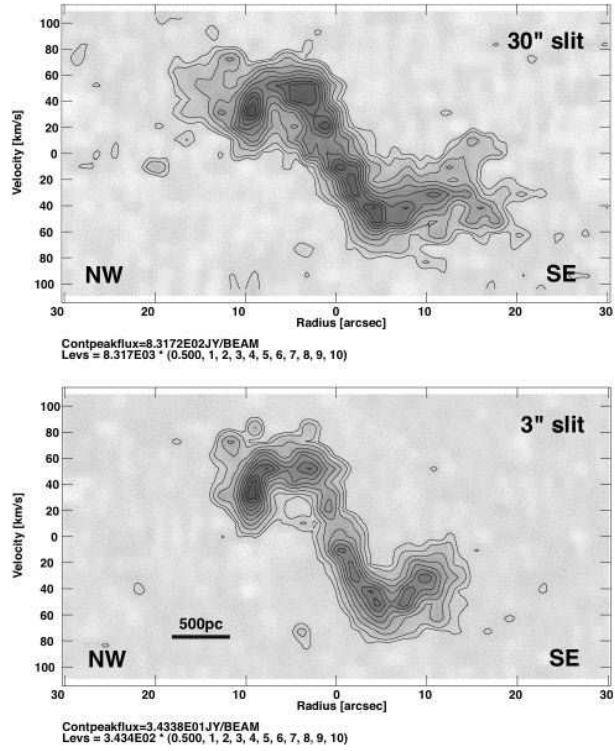}
\end{center}
\caption{
Position-velocity diagrams with $\rm PA=-44.45$ and slit widths of
$30\arcsec$ ({\it upper panel}) and $3\arcsec$ ({\it lower panel}).
The contours are 0.05, 0.1, 0.2, 0.3, 0.4, 0.5, 0.6, 0.7, 0.8, 0.9, 1.0
times the peak intensity. }
\label{fig:obspv}
\end{figure}

\begin{figure}
\begin{center}
\FigureFile(160mm,160mm){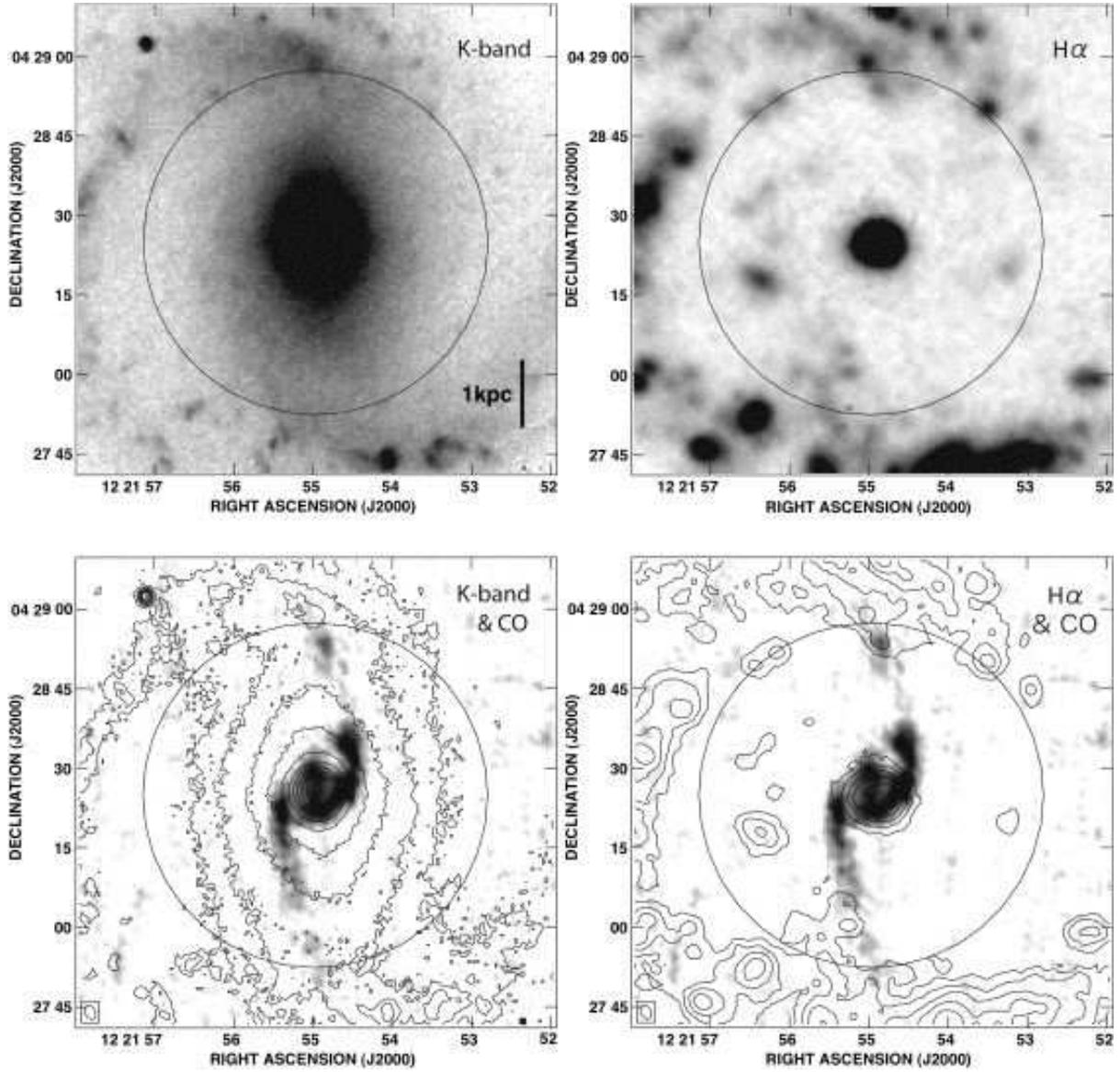}
\end{center}
\caption{
$K$-band and $\Ha$ images of the central $90\arcsec \times 90\arcsec$
region of NGC 4303. 
{\it Upper-left:} $K$-band image.
{\it Lower-left:} $K$-band contours on CO grayscale image.
Contours are 0.008, 0.012, 0.018, 0.027, 0.041, 0.061, 0.091, 0.14, 0.21,
0.31, 0.46, 0.69, 1.0 times the peak $K$-band intensity.
{\it Upper-right:} $\Ha$ image.
{\it Lower-right:} $\Ha$ contours on CO grayscale image.
Contours are 0.031, 0.063, 0.125, 0.25, 0.5, 1.0 times the peak $\Ha$
intensity.
The circles represent the field-of-view (FWHM) of CO observations.}
\label{fig:optimages2}
\end{figure}

\begin{figure}
\begin{center}
\FigureFile(160mm,160mm){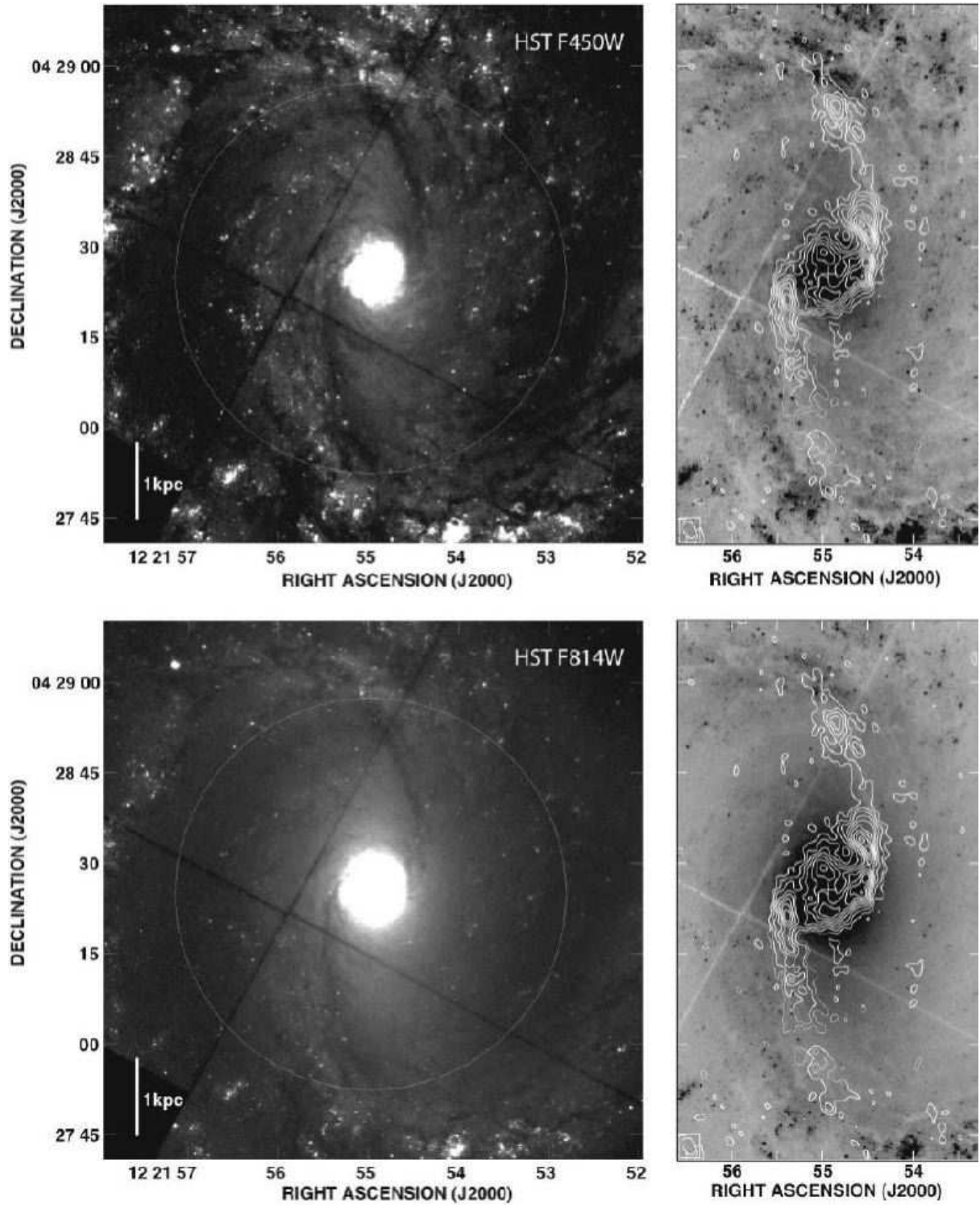}
\end{center}
\caption{HST images of the central $90\arcsec \times 90\arcsec$
region of NGC 4303, the same region covered in Figure \ref{fig:optimages}.
{\it Upper panels:} HST F450W band images.
{\it Lower panels:} HST F814W band images.
Right panels are the central vertical strips of the left images,
and CO contours are overlaid.
Contours are 0.1, 0.2, 0.3, 0.4, 0.5, 0.6, 0.7, 0.8,
0.9, 1.0 times the peak CO intensity.
The circles represent the FoV (FWHM) of CO observation.
The bright points across the images are very likely stellar clusters.
They are distributed predominantly at the leading (downstream) side of
the gas ridges, i.e. the eastern side of the southern ridge and 
the western side of the northern ridge.}
\label{fig:hstglobal}
\end{figure}

\begin{figure}
\begin{center}
\FigureFile(80mm,80mm){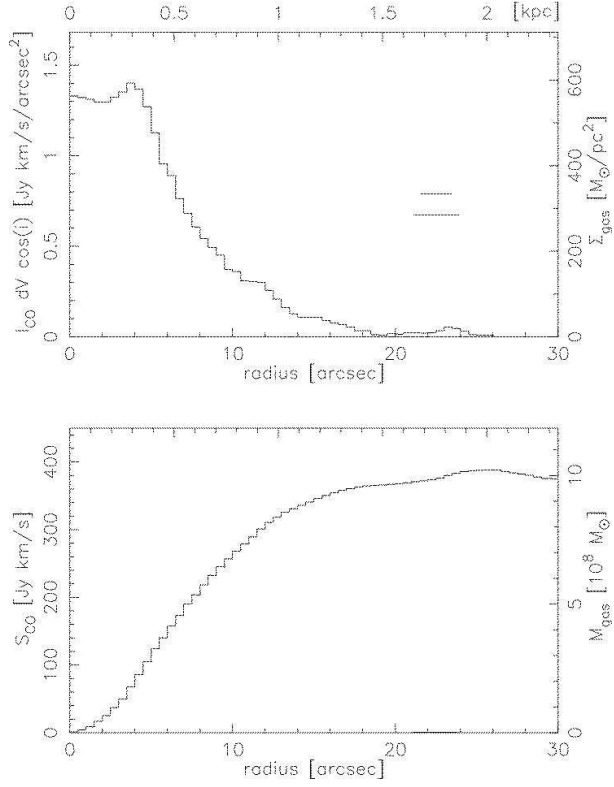}
\end{center}
\caption{
Radial profiles of the integrated-intensity (inclination corrected).
The physical scales at the top and right axes were calculated using
the distance of 16.1 Mpc and a conversion factor of
$X_{\rm CO}=1.8 \times 10^{20} \,{\rm cm^{-2} [K\,km\,s^{-1}]^{-1}}$.
We assumed $M_{\rm gas} = 1.41 M_{\rm H_2}$, taking He and the other
elements into account.
The two horizontal lines indicate the sizes of the major and minor axes
of the synthesized beam.}
\label{fig:prof}
\end{figure}

\begin{figure}
\begin{center}
\FigureFile(120mm,120mm){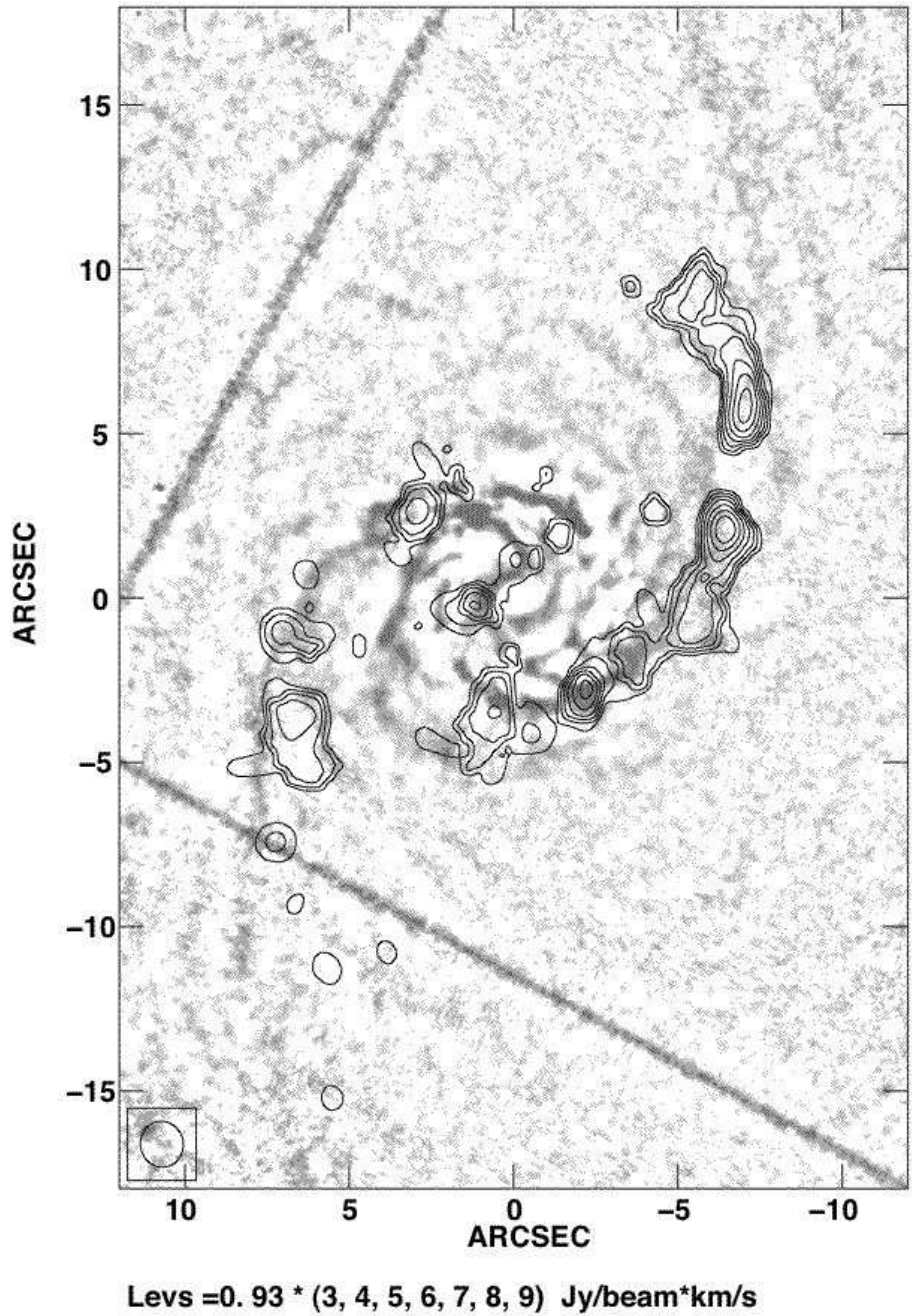}
\end{center}
\caption{High-resolution (uniform weighting) CO contours at the top of
an unsharp-masked $HST$ F814W image.
The contours are 0.3, 0.4, 0.5, 0.6, 0.7, 0.8, 0.9 times the peak
intensity, $0.9$ Jy/beam*km/s.}
 \label{fig:couni}
\end{figure}

Sharp velocity gradients exist across the offset ridges; in the
channel maps (Figure \ref{fig:chmaps}), the southern CO ridge shifts
east (toward left) as the velocity decreases, and the northern ridge
shifts west (right) as the velocity increases. The isovelocity
contours run along the offset ridges in the velocity field map
(Figure \ref{fig:comaps}).  The velocity differences across the
ridges, $\sim 4\arcsec$, is about $\sim 20\kmps$ (i.e. three channels)
in projection on the sky.

The typical brightness temperature is $T_{\rm b} \sim 1\K$ along the offset
ridges (Figure \ref{fig:chmaps}). Assuming an excitation temperature
$T_{\rm ex} = 10\K$ for optically-thick molecular gas \citep{sco87}
and the cosmic microwave background temperature of $T_{\rm CMB}= 2.7\K$,
the area filling factor within the beam ($2.4\arcsec\sim 200\pc$) is
$f\sim 0.1$ .
Thus, the emission arises from unresolved clumpy structures,
presumably molecular clouds.
The average gas surface density in the beam is typically
$\Sigma_{\rm gas}=150 \Msunpcpc$. Therefore, molecular clouds are
dense with an average surface density of 
$\Sigma_{\rm MC}\sim \Sigma_{\rm gas}/f\sim 1000\Msunpcpc$.
This is much denser than $170\Msunpcpc$ for clouds
in the Galactic disk \citep{sol87}, but less dense than $2500\Msunpcpc$
for clouds in the Galactic center \citep{oka01}.
Note that a velocity dilution might underestimate the value $f$ by a
factor of 2, although it does not affect the following discussion.
The typical velocity width of the offset ridges is about $20\kmps$
(Figure \ref{fig:chmaps}), and that of molecular clouds in the Galaxy
is $9\kmps$ (FWHM; \cite{sco87}).

\subsection{Star Formation around Offset Ridges}\label{sec:ressf}


Evidence of star formation (SF) around the offset ridges is inferred
from many bright points in the {\it HST} images (Figure \ref{fig:hstglobal}),
although
active SF is not obvious in the $\Ha$ image (Figure \ref{fig:optimages2}
{\it right}).
Each of the bright points is
very likely a stellar cluster.
Their luminosities indicate that the
masses are $10^{3-4}\Msun$. Assuming an extinction $A_V=1\,\rm mag$
and a color excess $0.71$ \citep{car89}, the typical intrinsic color in
F450W-F814W is as blue as $-0.7 \rm \, mag$ (AB magnitude),
indicating their young ages, $\lesssim 10^7\yr$ (see Appendix).
Because of the large scatter in color ($\pm 0.5 \rm\, mag$),
we could not detect any significant systematic color change
as a function of the distance from the ridges.

The clusters are distributed predominantly at the leading side of the gas
ridges, i.e. at the downstream side assuming that the corotation
radius is around the bar end.
No obvious cluster is found at
the opposite (upstream) side, which is especially notable
for the southern ridge. Therefore, the star formation was initiated
during, or after, passage of the parent molecular cloud across the offset ridges.
Some clusters are found {\it far} from the gas ridges.
The rotation timescale around the galaxy, $10^8\yr$,
is ten-times longer than the cluster age, $\lesssim 10^7\yr$.
Thus, those clusters should have been formed not {\it on} the ridges,
but {\it long after} passage of the cloud across the ridges.

The conversion of molecular gas to stars in a molecular cloud is normally
a very inefficient process in the Galaxy with a conversion efficiency of
0.01-0.1 \citep{eva91,lad92}. Consequently, the parent molecular clouds
of the stellar
clusters should have had masses of $10^{4-6}\Msun$, which are typical
for Galactic molecular clouds \citep{sco87}.

\subsection{Comparison with Other CO Maps}

Three CO maps of NGC 4303
have been published from mm-wave interferometer observations,
at the Owens Valley Radio Observatory (OVRO; \cite{sch02}), at
the Nobeyama Millimeter Array (NMA; \cite{sof03}, and this paper),
and at the Berkeley-Illinois-Maryland Association (BIMA; \cite{hel03}).

The BIMA map has three-times lower sensitivity of $\sim 47\, \rm mJy$
and resolution of $\sim 6\arcsec$, but includes zero-spacing baselines
(extended components) from single dish observations. It is thus good
for tracing the large-scale distribution of molecular gas.
The BIMA map shows significant CO emission along the outer spiral arms.
The brightness is comparable between the offset ridges and
spiral arms, though star formation is more active in the arms
(Figure \ref{fig:optimages}).

The OVRO map has similar sensitivity ($\sim 20 \, \rm mJy$) and
resolution ($\sim 2\arcsec$) to our map. The CO distribution and velocity
structure are consistent between the OVRO and NMA maps. The recovered flux
compared with the single-dish data \citep{ken88}, however, is 2-times
higher in NMA than in OVRO.
We compared the two data cubes\footnote{OVRO data cube was kindly provided
by Eva Schinnerer.}, and found that the NMA flux is $\sim 1.4$ times
larger in spectrum and $\sim 2$ times greater in integrated intensity.
The lower OVRO flux probably results from a combination of the
assumed calibrator flux ($\sim 10$ Jy vs $\sim 14.5$ Jy) and
a lack of short-spacing {\it uv} coverage in the OVRO data.
This difference does not affect the discussions made in this
paper and in \citet{sch02}.


\section{Gas Dynamics in NGC 4303}\label{sec:mod}
\begin{figure}
\begin{center}
\FigureFile(160mm,160mm){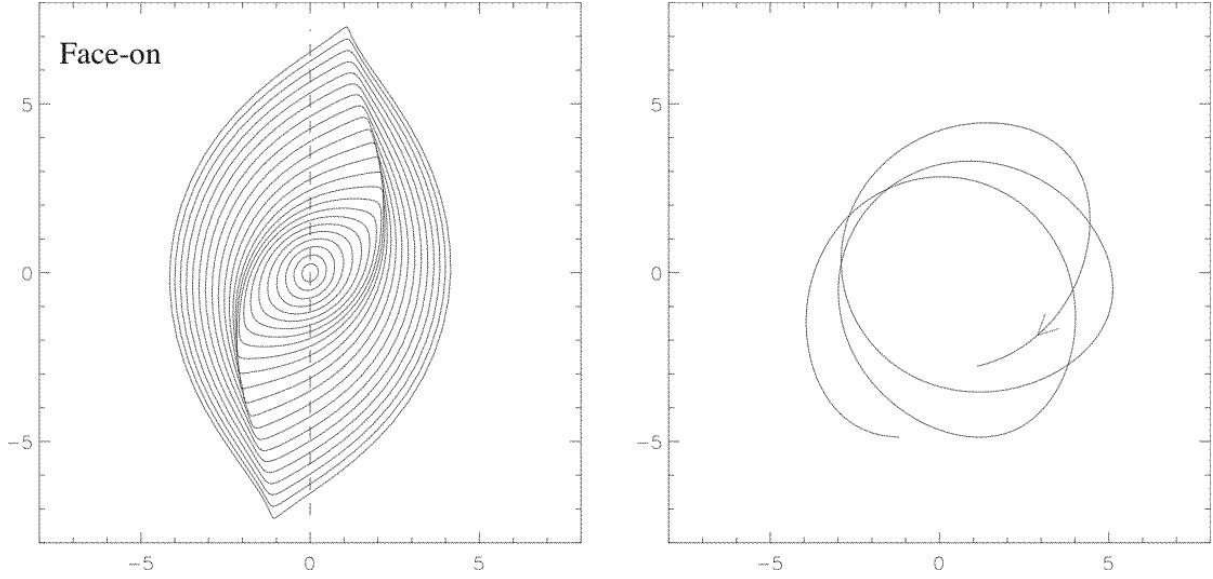}
\end{center}
\caption{
{\it Left:} Gas orbits in a bar in the rotating frame with the bar (clockwise
rotation).
The dashed line indicates the bar direction. {\it Right:} An orbit in the
non-rotating reference frame. The gas is rotating clockwise in both panels.
We used the potential
$\Phi(R, \varphi) = (1-\epsilon \cos 2 \varphi) \frac{v_0^2}{2} \log(1+(R/a)^2)$, where $a=1$, $v_0=1$ and $\epsilon=0.04$. The pattern speed of the bar is
$\Omegab=0.1$. The two inner Lindblad resonances and the corotation radii are
$R_{\rm ILR}=1.1, 2.0$, and $R_{\rm CR}=9.9$, respectively.
Assuming a bar length of 6 kpc ($77\arcsec$) and rotational velocity
of $160\kmps$ at the radius of 3 kpc for NGC 4303,
the units of the model become $a=300\pc$ ($3\arcsec.8$), $v_0=160\kmps$, and
$\Omegab=53\kmps{\rm \,kpc^{-1}}$.}
 \label{fig:modorb}
\end{figure}

We use a cloud-orbit model to interpret the gas dynamics in NGC 4303.
To derive the stellar orbits in the bar, the equations of motion for a test particle
are solved in a bar potential \citep{con80,bin87}. \citet{wad94}
included the damping force term, ``$-2 \lambda \dot{R}$'', in the
equations of motion, the force negatively proportional to
the velocity $\dot{R}$, and obtained the orbits of collisional gas clouds.
This model has provided successful
interpretations for observed barred and spiral galaxies \citep{sak00,
kod02,ono04}.
Figure \ref{fig:modorb}, \ref{fig:modmaps}, and \ref{fig:modpv} show
the model gas orbits, density and velocity field maps, and position-velocity
diagrams. The gas and bar are rotating clockwise.
The model is scalable; it fits to NGC 4303 when the units are $300\pc$
in length and $160\kmps$ in velocity. The pattern speed and bar radius
(corotation radius) are scaled to $53\kmps\,\kpc^{-1}$ and $3 \kpc$,
respectively, which is consistent with the bar radius
in Figure \ref{fig:optimages}.

\begin{center}
------ Figure \ref{fig:modorb}  ------ 
\end{center}

\subsection{Offset Ridges and Central Spirals}

\begin{figure}
\begin{center}
\FigureFile(160mm,160mm){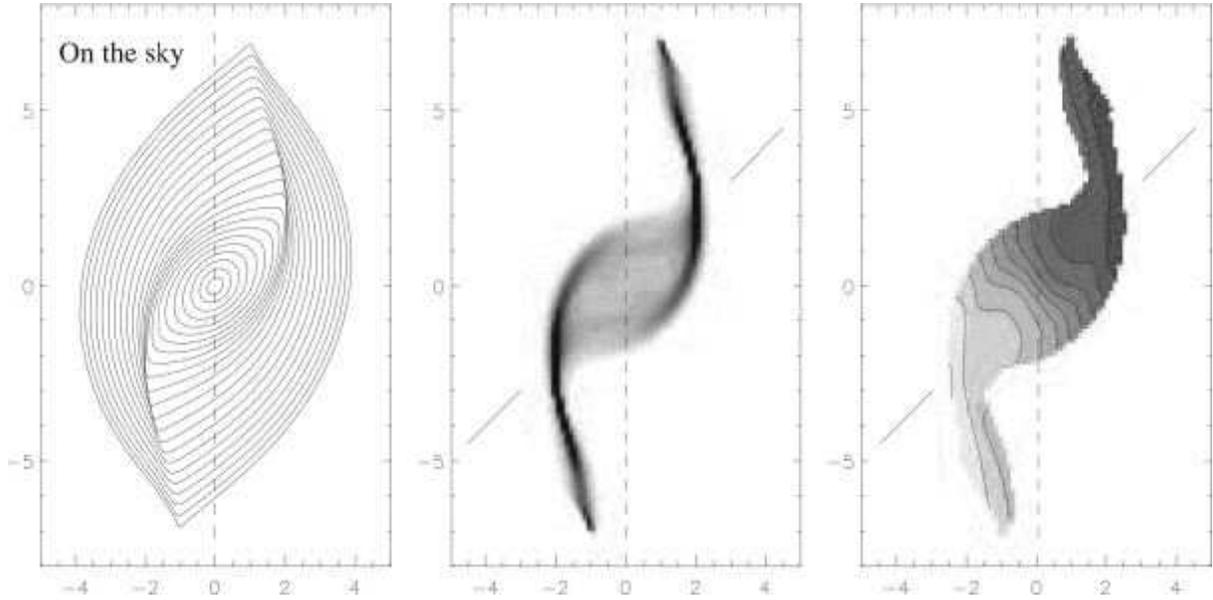}
\end{center}
\caption{
{\it Left:} Gas orbits in a bar projected on the sky. The position
angle and inclination are $-45\degree$ and $30\degree$, respectively.
We first rotated the gas orbits (Figure \ref{fig:modorb} {\it left})
counterclockwise by $4\degree$, so that the bar (dashed line) runs
vertically, like that in NGC 4303.
{\it Middle:} Density map, calculated from the gas orbits based
on the speed of gas motions. The radial density profile is set to
be $\exp(-r/4)$ for $r>2$ and constant for $r\le2$. We do not show
low density regions. The solid lines indicate the major axis
(${\rm P. A.}=-45\degree$).
{\it Right: } Velocity field, calculated from the gas orbit model.}
\label{fig:modmaps}
\end{figure}

Figure \ref{fig:modmaps} shows the projection of the gas orbits with
a position angle of $-45\arcdeg$ and an inclination of $30\arcdeg$
(Table \ref{tab:kinparm}).
The stellar bar runs vertically. 
The density map ({\it middle}) reproduces the observed offset ridges
and the central spiral arms in Figure \ref{fig:comaps}.
To calculate the density, we used the timescale of
passage along the orbit as a weighting function.  The offset ridges and
central spiral arms appear to be dense for two reasons:
(1) the gas orbits are crowded at the leading side of the bar, and
(2) passing the apocenter of an orbit takes longer time than passing
the pericenter.

There is a common expectation that the offset ridges are produced by
hydrodynamic shock; the tangential orbital velocity of
the gas is damped by the shock, and the gas falls toward the galactic
center along the offset ridges.
The orbit model, however, indicates that {\it no shock is required to
form the dense and narrow offset ridges.}
In fact, the sharp turns are apparent in the {\it rotating}
frame with the bar (Figure \ref{fig:modorb} {\it left}), but disappear
in the rest frame ({\it right}). The tangential velocity becomes
very close to the rotation speed of the bar on the offset ridges, which
produces the apparent turns.
This fact is particularly important when the orbits of young stars
born recently in molecular clouds are discussed.
We note that this argument does not exclude the possibility of shock as
a secondary effect.

\begin{center}
------ Figure \ref{fig:modmaps}  ------ 
\end{center}

The velocity field (Figure \ref{fig:modmaps} {\it right}) is consistent
with the observations (Figure \ref{fig:comaps} {\it middle}).
The major axes of elongated gas orbits are aligned with that of the
galaxy at the central disk, and therefore, the isovelocity contours
are similar to those of pure circular rotation (spider diagram).
Slight deviations from pure circular rotation are apparent
along the spiral arms on the central disk; we found similar
deviations in the observed velocity field (Figure \ref{fig:comaps}
{\it middle}).
At the outer part, the isovelocity contours run along the offset ridges
as found in the observations. The contours parallel to the offset ridges
indicate abrupt velocity changes across the ridges, although no shock
is associated.

\begin{figure}
\begin{center}
\FigureFile(80mm,80mm){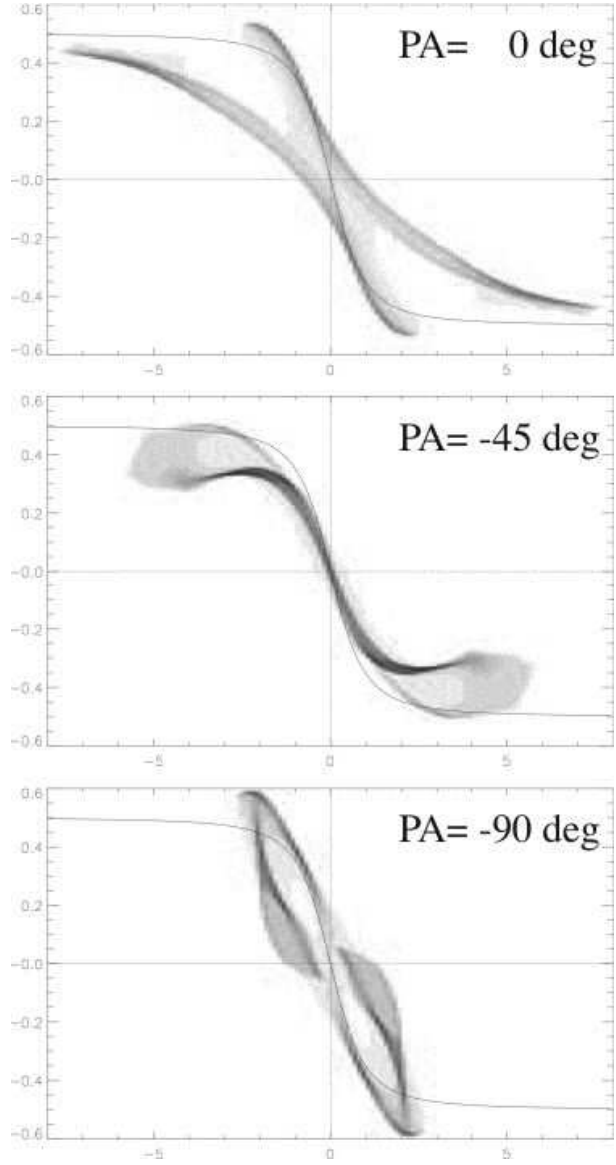}
\end{center}
\caption{Position-velocity diagrams of the model in Figure \ref{fig:modmaps}
observed from the three different position angles, i.e. $0\degree$ ({\it top}),
$-45\degree$ ({\it middle}), and $-90\degree$ ({\it bottom}).
The diagram with PA$=-45\arcdeg$ corresponds
to NGC 4303. The same gas distribution shows quite different patterns;
our determination of PA fits the data.
The solid curves are the rotation curve given by the assumed potential.}
\label{fig:modpv}
\end{figure}

\begin{center}
------ Figure \ref{fig:modpv}  ------ 
\end{center}

Figure \ref{fig:modpv} shows position-velocity (PV) diagrams 
with three different position angles (PA). The diagram with
${\rm PA} = -45\arcdeg$ corresponds to NGC 4303 and is similar to
the observed one. The model reproduced the observations in
space (maps) and velocity (PV diagram), indicating that the orbit
model represents the gas motions in NGC 4303.
The other two diagrams, i.e. ${\rm PA} = 0\arcdeg$ and $-90\arcdeg$,
are projections from the edge-on and pole-on directions of the bar,
respectively. The same gas distribution shows quite different patterns;
our determination of the PA and
inclination in \S \ref{sec:rescent} were suitable for NGC 4303.

\subsection{Cloud-cloud Collisions as a Damping Force}\label{sec:damping}

The damping force term ``$-2 \lambda \dot{R}$'' provides the damping
timescale, $t_{\rm damp}=1/\lambda$, in which random motions decay.
For the above model, we defined the damping coefficient
$\Lambda \equiv \lambda/\kappa_0 =0.1$
as in previous studies \citep{wad94,sak00}. The orbits are
insensitive to the change of $\Lambda$ by a factor of a few \citep{wad94}.
We discuss its physical cause in an order-of-magnitude approximation.
Assuming a constant rotation velocity $V=160\kmps$ at a radius of $1\kpc$
as for NGC 4303, the epicyclic frequency becomes
$\kappa_0 \sim 200\kmps\,\kpc^{-1}$. The damping timescale is then
$t_{\rm damp} = 1/\Lambda \kappa_0 \sim 5\times 10^7 {\rm \, yr}$
for $\Lambda=0.1$.

We have shown the presence of dense molecular clouds in NGC 4303
(\S \ref{sec:resoff}). The damping timescale is close to the
collisional timescale of clouds calculated as follows.
NGC 4303 has a gas mass of $10^8\Msun$ between radii
of $1-2\kpc$ (Figure \ref{fig:prof}).
Assuming a cloud mass $M_{\rm MC}=10^5\Msun$ and diameter $D=20\pc$,
the surface number density of molecular clouds is
$N_{\rm MC}=100 \kpc^{-2}$.
If the cloud-cloud velocity dispersion is $\sigma=10\kmps$
and the galactic molecular disk is thin (2-D) in the central region,
the collisional timescale
between molecular clouds becomes $t_{col}=1/N_{\rm MC} D \sigma \sim 4\times
10^7 {\rm \, yr}$. Close encounters would occur on a slightly shorter timescale
but on the order of $10^7$\yr. These are close to the damping timescale
for $\Lambda \sim 0.1$.
Cloud collisions could be the cause of the damping force.

Ram pressure from ambient gas is negligible.
Using the density of the ambient gas $\rho$, the relative velocity between
a cloud and the ambient gas $v$, the cloud cross section $A$, and
the mass $M_{\rm MC}$, the acceleration on the cloud is $a_{\rm ram}=
\rho v^2 A/M_{\rm MC}$. The velocity difference would decay in
$t_{\rm ram}=v/a_{\rm ram}$. If the ambient gas is atomic hydrogen
with a density of $1 {\rm \, cm^{-3}}$ and a velocity of $v=10\kmps$,
$t_{\rm ram}\sim 10^9 {\rm \, yr}$, which is much longer than $t_{\rm col}$.

\subsection{Shear Velocity and the Survival of Molecular Clouds}\label{sec:esc}

\citet{dow96} discussed that molecular clouds might be destroyed by
strong shear around offset ridges. Assuming that the shapes and
orientations of the gas orbits in Figure \ref{fig:modmaps} {\it left}
are all correct, molecular clouds have only the velocity components
tangential to our line-of-sight before the ridges, and only the
line-of-sight velocity components after the ridges.
The velocity difference across the ridges
is then $\sim 40\kmps$ from the observed velocity difference of $20\kmps$ and
an inclination of $30\arcdeg$. The width of the ridge is $\sim 300\pc$
($\sim 4\arcsec$), and thus the velocity gradient (shear) is about
$0.13\kmps\,\pc^{-1}$.

The surface density of a molecular cloud in NGC 4303 is about
$\Sigma_{\rm MC}=1000\Msun\,\pc^{-2}$ (\S \ref{sec:resoff}). Assuming a
cloud radius $R=10\pc$, the escape velocity from the cloud,
$v_{\rm esc}= (2 \pi G \Sigma_{\rm MC} R)^{1/2}$, is $16\kmps$. Thus,
a sharp velocity gradient of $1.6\kmps\,\pc^{-1}$ is necessary to destroy
the molecular cloud. This is an order of magnitude above the observed
shear velocity gradient in NGC 4303. Even for some molecular clouds
with lower surface densities of $100\Msun\,\pc^{-2}$ (typical for the Galactic
molecular clouds), the escape velocity
is still higher than the observed shear velocity.
Thus, shear is unlikely to destroy molecular clouds.
{\it The clouds are likely to survive passage across the ridges.}

\section{Collision-Induced Star Formation in the Bar}\label{sec:sf}

The existence of molecular clouds is necessary but not sufficient for
star formation \citep{moo88,sco89,fuk99}. There is considerable evidence
that cloud-cloud collisions may initiate star formation
\citep{lor76,sco86,ode92,has94}. The efficiency of collisional dissipation
(\S\ref{sec:damping}) also supports the cloud-based mechanism for triggering
star formation. We explore the cloud-cloud collision (or close encounter)
model to account for star formation.

\subsection{Star Formation at the Leading Side of Ridges}
The damping timescale of random motions is close to the rotation
timescale of $10^8\yr$ around the galactic center (\S \ref{sec:damping}).
Thus, the dissipation is not very effective, and newborn stars should
follow similar orbits to those of the parent molecular clouds (Figure
\ref{fig:modorb}) within about one rotation timescale.
Comparing the distribution of the young clusters
(Figure \ref{fig:hstglobal}) to the orbits (Figure \ref{fig:modmaps}
{\it left}), we find that the majority of the clusters are aligned
on the outermost orbits in Figure \ref{fig:modmaps}.
Hence, most star formation occurs at the outer
ends of the offset ridges. The same consideration can
be applied to other barred galaxies (see \cite{she02});
the outer ends of offset ridges are the primary loci of star formation
in the bars.

\begin{figure}
\begin{center}
\FigureFile(80mm,80mm){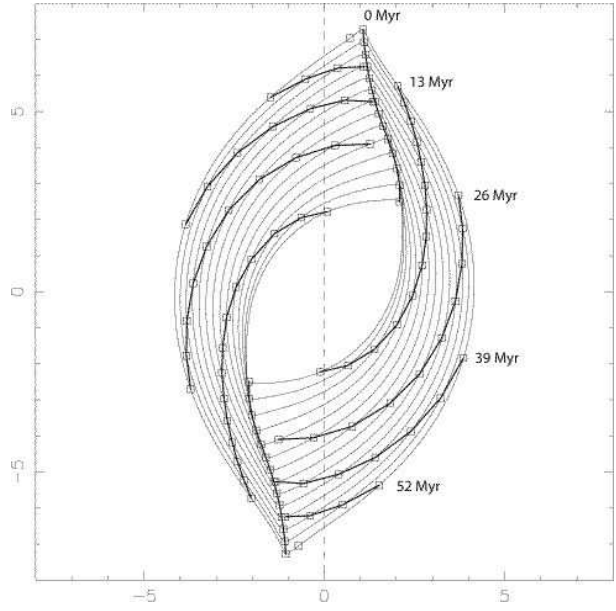}
\end{center}
\caption{Isochrones of gas flows. The thick solid lines indicate 0, 13, 26, 39, 52 Myr after passage over the turning points (the offset ridges). Gas orbits (same as Figure \ref{fig:modorb}) are also shown for a reference.}
\label{fig:modiso}
\end{figure}

We discussed in \S \ref{sec:ressf} that some of the stellar clusters have
large offsets from the gas ridges (Figure \ref{fig:hstglobal}).
Considering the isochrones along the orbits (Figure \ref{fig:modiso})
and the young cluster ages $\lesssim 10^7\yr$, they should have been
born long ($10^7\yr$) after the parent clouds cross the ridges.
Star formation is, therefore, triggered even after (but not before)
the offset ridges. This cannot be explained by the star formation
induced by galactic shocks. In Figure \ref{fig:modorb} {\it left},
the separation between orbits is remarkably narrow at the leading
side of the ridges.
The number density of clouds would
therefore be larger at the leading side, and possibly result
in more frequent cloud collisions (or close encounters).
The collision model explains the star formation after the ridges.

\begin{center}
------ Figure \ref{fig:modiso}  ------ 
\end{center}

The presence of HII regions at the leading side of the ridges has been
a problem for hydrodynamic models \citep{she02}.
They predict that the gas looses
the tangential velocity at offset shocks, and falls
toward the galactic center along the offset ridges
\citep[see their Figure 9]{reg99}.
If star formation occurs due to the shocks, i.e. after the tangential
velocity is damped, the newborn stars should also fall {\it along}
the offset ridges and cannot exist at their leading side (\cite{she02}).
The orbit model, however, naturally explains the co-motion of molecular
clouds and stars, and the star-forming regions at the leading side.

Young stars are bright in UV emission for $10^8\yr$ after birth
\citep{igl04}, which is close to the rotation timescale.
If the young stars follow the gas orbits for a single rotation
timescale and if the outer edges of the ridges are the primary
loci of star formation, the outermost orbits in Figure
\ref{fig:modmaps} {\it left} would stand out in UV emission from
young stars. This is the case for
NGC 4303 in the GALEX/UV image (Gil de Paz et al. 2006 in preparation).

\section{Cloud Model vs Hydrodynamic Model}

Most hydrodynamic models assume an isothermal gas with a sound
speed of $\sim 10\kmps$, which is similar to the observed
cloud-cloud velocity dispersions.
The corresponding effective temperature is about $10^4\K$, far higher
than the typical kinetic temperature of molecular gas ($\sim 10\K$).
Therefore, the hydrodynamic models represent not the diffuse
molecular gas, but an ensemble of a significant number of molecular clouds.
The difference between the two models arises from the number of
clouds being considered. Cloud-based dynamics assumes fewer clouds,
and thus, would be less dissipative than smooth hydrodynamic models.
The efficiency of
gas fueling to galactic centers could be lower than
what we expect from hydrodynamic simulations.

Obviously, there is diffuse, presumably continuous, gas around the
discrete clouds. The gas is cycling between the two phases due
to cloud formation, stellar feedback, ram pressure stripping of
the cloud envelope, etc. This cycling would affect the gas dynamics.
To discuss these effects, molecular
clouds have to be resolved in galaxies with higher
resolution observations (e.g. ALMA) and simulations (e.g. \cite{wad01}).

\section{Conclusions}\label{sec:con}

We observed the central 5 kpc of the barred galaxy NGC 4303 in the
CO(1-0) line with the Nobeyama Millimeter Array. We discussed the
gas distribution and dynamics, and compared them with
young stellar cluster distribution in {\it HST} images.
The main conclusions are:

1.
NGC 4303 has a central gas concentration (disk) and offset ridges
along its stellar bar. The central disk follows pure circular rotation
with slight perturbations due to spiral arms on the disk. Sharp velocity
gradients are observed across the offset ridges.

2.
The brightness temperature in the $2\arcsec.4$-resolution map is much lower
than the typical excitation temperature of molecular gas in the Galaxy.
Assuming that the gas is optically thick in CO(1-0), the beam-filling
factor is about 0.1, indicating the existence of dense molecular clouds
($\sim 1000\Msun \, \pc^{-2}$) in the offset ridges.

3.
The velocity gradient across the offset ridges is about
$0.13\kmps \,\pc^{-1}$. This is an order of magnitude smaller than
the gradient necessary to break up existing molecular clouds.
Therefore, the clouds survive after passing the offset ridges.

4.
Many young stellar clusters are found in {\it HST} images, although
star formation is not apparent in the $\Ha$ image.
The clusters have ages of $\lesssim 10^7\yr$ and masses $10^{3-4}\Msun$.
If the star-formation efficiency is 0.01-0.1,
their parent molecular clouds should have had masses of $10^{4-6}\Msun$.

5.
The young clusters are distributed predominantly at the downstream
side of the offset ridges. Some of them are located far from the
offset ridges and, therefore, are born long after their parent
clouds pass the ridges.

6.
We used a cloud orbit model to simulate the gas dynamics.
The dynamics due to slightly-collisional clouds reproduces
the central gas disk,
narrow offset ridges, and their associated velocity structure.
The offset ridges are not necessarily associated with shock.

7.
The strength of viscous (damping) force is a parameter of choice in
the cloud orbit model. The best-fit value gives a timescale of $5 \times
10^7\yr$ for the decay of random motions, which is close to the
timescale of cloud-cloud collisions or close interactions in NGC 4303.

8.
Cloud collisions naturally explain the formation of the young
clusters far from the offset ridges. Cloud-orbits are densely
populated at the leading side of the ridges, indicating a high
collision frequency and star formation.

\appendix

\section{Color Evolution of a Stellar Cluster}\label{sec:col}

\begin{figure}
\begin{center}
\FigureFile(80mm,80mm){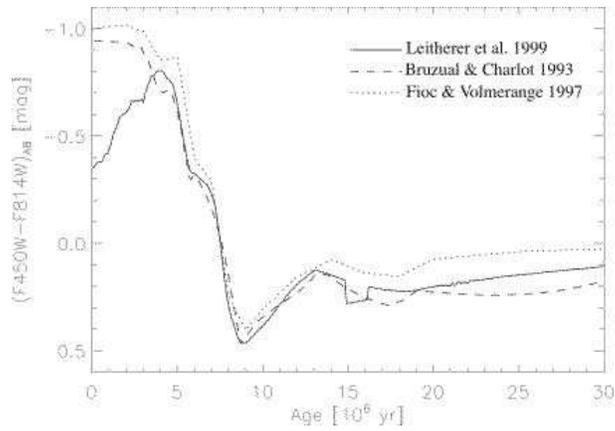}
\end{center}
\caption{Model color evolution of stellar clusters.}
\label{fig:color}
\end{figure}

Figure \ref{fig:color} shows the model color evolution of a stellar
cluster. Three photometric synthesis models are compared:
Bruzual and Charlot (1993, hereafter, BC93), Fioc and Rocca-Volmerange
(1999, Pegase), and Leitherer et al. (1999, Starburst99).
We used the Salpeter initial mass function (IMF) with
a mass range $0.1-100\Msun$ for BC93, the Scalo98 IMF with
a range $0.1-120\Msun$ for Pegase, and the IMF having indices
of 1.3 for $0.1-0.5\Msun$ and 2.3 for $0.5-100\Msun$ for Starburst99.
The F450W-F814W color drops sharply around $7\times 10^6\yr$ for all
three models. The uncertainties are a factor of 2 in time and
$0.1\,\rm mag$ in magnitude. The stellar clusters bluer than
$+0.0\,\rm mag$ (\S \ref{sec:resoff}) are, therefore, younger than
$\lesssim 10^7\yr$ (and $2\times 10^7\yr$ at the oldest).

\begin{center}
------ Figure \ref{fig:color} ------
\end{center}

\newpage

We thank Tsutomu Takamiya, Kotaro Kohno, Makoto Hidaka, Hiroyuki Nakanishi,
and Sachiko Onodera for useful discussions and collaborations
in the Virgo CO survey project. We also thank the NMA staff for their help
with observations, and Keiichi Wada for fruitful discussions.
J. K. thanks Ryo Kandori for helping with the data analysis,
Eva Schinnerer for providing her CO data,
Armando Gil de Paz and Samuel Boissier for providing their paper prior to
publication, Nick Scoville and Jenny Patience for carefully reading
the manuscript, and
Kartik Sheth for useful comments. We thank an anonymous referee for
comments, which helped to clarify the paper.
J. K. was financially supported by the Japan Society
for the Promotion of Science for Young Scientists. This work was partially
supported by National Science Foundation under grant AST-9981546.


\newpage
\draft

\clearpage
\begin{table}
  \caption{Parameters of NGC 4303.}\label{tab:gnlparam}
  \begin{center}
    \begin{tabular}{lrc}
      \hline\hline
      Parameter                       & Value & Reference \\
      \hline
      Alias                           & M61      &   \\
      Hubble type                     & SAB(rs)bc& 1 \\
      Nuclear activity                & H        & 2 \\
      Distance (Mpc)                  & 16.1     & 3 \\
      Linear scale of $1\arcsec$ (pc) & 78.1       & \\
      P.A.(isophotal)                 & & \\
      Inclination (isophotal)         & & \\
      $\vsys$ ($\kmps$)               & 1569     & 1 \\
      $D_{25}^0$ (arcmin)             & 6.46     & 1 \\
      $B_T^0$ (mag)                   & 10.12    & 1 \\
      $S_{CO(1-0)}(45\arcsec)$ ($\jkmps$) & $494 \pm 87$ & 4 \\
      $S_{CO(1-0)}(13\arcsec)$ ($\jkmps$) & $193 \pm  6$ & 5 \\
      \hline
    \end{tabular}
  \end{center}
References:
(1) \cite{dev91};
(2) \cite{ho97};
(3) \cite{fer96};
(4) \cite{ken88}, $45\arcsec$ (FHWM) beam centered at
$\alpha_{1950}=12^{\rm h}19^{\rm m}21^{\rm s}.4$ and
$\delta_{1950}=+4\arcdeg44\arcmin58\arcsec.0$;
(5) \cite{nis01}, $16\arcsec$ (FHWM) beam centered at
$\alpha_{1950}=12^{\rm h}19^{\rm m}21^{\rm s}.4$ and
$\delta_{1950}=+4\arcdeg44\arcmin58\arcsec.0$
\end{table}

\begin{table}
  \caption{Observational parameters.}\label{tab:obsparm}
  \begin{center}
    \begin{tabular}{lc}
      \hline\hline
      Parameter     & Value \\
      \hline
      Year          & 1999 Dec.  - 2000 Feb.  \\
      Field center  &                         \\
      \,\,\,\,\,\,$\alpha_{1950}$, $\delta_{1950}$ & $12^{\rm h}19^{\rm m}21^{\rm s}.60$, $+4\arcdeg45\arcmin03\arcsec.0$  \\
      \,\,\,\,\,\,$\alpha_{2000}$, $\delta_{2000}$ & $12^{\rm h}21^{\rm m}54^{\rm s}.97$, $+4\arcdeg28\arcmin24\arcsec.9$  \\
      Field of view & $65\arcsec$ \\
      Array configuration & AB, C, D \\
      Observing frequency & $114.65925\GHz$   \\
      Band width          & $512\MHz$ \\
      \hline
    \end{tabular}
  \end{center}
\end{table}

\begin{table}
  \caption{Parameters of CO(1-0) cubes.}\label{tab:cbparam}
  \begin{center}
    \begin{tabular}{lcc}
      \hline \hline
      Parameter                    & Value  & Value \\ 
      \hline
      Configuration                & AB+C+D & AB+C+D \\
      Weighting                    & NA     & UN \\
      Synthesized beam             & $2\arcsec.8  \times 1\arcsec$.9,$166\arcdeg$   &  $1\arcsec.4  \times 1\arcsec$.3,$11\arcdeg$   \\
      $\Delta V$ ($\kmps$)         & 10.4 & 31.2 \\
      rms ($\rm mJy \, beam^{-1}$) & 21 & 24 \\
      $T_b$ for $\rm 1 \,Jy \, beam^{-1}$ & 17.3 & 50.5\\
      $T_b$ for $\rm 1 \,Jy \, arcsec^{-2}$ & 91.9 & 91.9\\ 
      \hline
    \end{tabular}
  \end{center}
\end{table}

\begin{table}
  \caption{Kinematic parameters of the nuclear disk}\label{tab:kinparm}
  \begin{center}
    \begin{tabular}{lc}
      \hline\hline
      Parameter     & Value \\
      \hline
      Dynamical center  &                         \\
      \,\,\,\,\,\,$\alpha_{1950}$, $\delta_{1950}$ & $12^{\rm h}19^{\rm m}21^{\rm s}.67$, $+4\arcdeg45\arcmin03\arcsec.7$  \\
      \,\,\,\,\,\,$\alpha_{2000}$, $\delta_{2000}$ & $12^{\rm h}21^{\rm m}54^{\rm s}.94$, $+4\arcdeg28\arcmin25\arcsec.6$  \\
      $V_{sys}$ ($\kmps$) & 1556.5 \\
      P.A. (deg)  & -44.45 \\
      $i$  (deg)  & 29.13\\
      \hline
    \end{tabular}
  \end{center}
\end{table}

\end{document}